\documentclass[aps,prl,floats,amsmath,showpacs,twocolumn]{revtex4}
\usepackage{graphicx}

\begin{document}

\title{Unusual structural tuning of magnetism in cuprate perovskites}

\author{Jorge \'I\~niguez$^{1,2}$ and Taner Yildirim$^1$}

\affiliation{$^{1}$NIST Center for Neutron Research, National
Institute of Standards and Technology, Gaithersburg, MD 20899\\
$^{2}$Dept. of Materials Science and Engineering, University of
Maryland, College Park, MD 20742}

\begin{abstract}
Understanding the structural underpinnings of magnetism is of great
fundamental and practical interest. Se$_{1-x}$Te$_x$CuO$_3$ alloys are
model systems for the study of this question, as composition-induced
structural changes control their magnetic interactions. Our work
reveals that this structural tuning is associated with the position of
the supposedly {\sl dummy} atoms Se and Te relative to the
super-exchange (SE) Cu--O--Cu paths, and not with the SE angles as
previously thought. We use density functional theory, tight-binding,
and exact diagonalization methods to unveil the cause of this
surprising effect and hint at new ways of engineering magnetic
interactions in solids.
\end{abstract}

\pacs{75.10.-b,75.30.Et,71.27.+a,75.80.+q}

\maketitle

The 3$d$ transition metal oxides are very important materials for they
have been a source of novel and intriguing physical phenomena such as
high-$T_c$ superconductivity, colossal magneto-resistance, and
magneto-electricity. Not surprisingly, a lot of effort is being devoted
to understanding the microscopic interactions that determine the
behavior of these systems. Here we are concerned with a particularly
important topic, namely, the structural dependence of the magnetic
couplings relevant in insulators (e.g., direct- and
super-exchange). This question already received a lot of attention in
early studies of magnetism in solids~\cite{and63}, and a renewed
interest in it is being driven by current intense work on
magneto-electric materials.

\begin{figure}[b!]
\begin{center}\includegraphics[width=7cm]{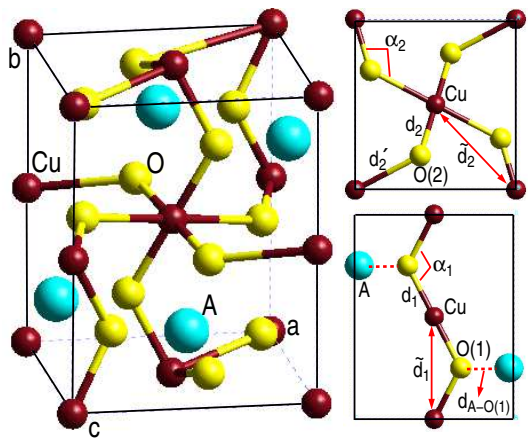}\end{center}
\caption{Left: Unit cell of ACuO$_3$ (A=Se,Te). Right top:
symmetry-equivalent Cu--O(2)--Cu groups in $ac$ plane. Right bottom:
Cu--O(1)--Cu chains along $b$ direction. Note there are two types of
oxygens in the unit cell. Relevant structural parameters are defined.}
\label{fig1}
\end{figure}

\begin{figure}
\includegraphics[width=8cm]{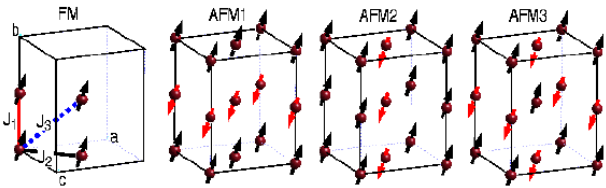}
\caption{Spin structures considered in this work. In the ``FM'' panel,
only the four Cu atoms in the unit cell are shown, and the exchange
constants are defined.}
\label{fig2}
\end{figure}

Se$_{1-x}$Te$_x$CuO$_3$ alloys (STCO)~\cite{sub99,law03} are model
systems for the study of these issues. They crystallize in a perovskite
structure that is strongly distorted because both Te$^{+4}$ and
Se$^{+4}$ are relatively small. Increasing $x$ results in structural
distortions that, in turn, switch the magnetic ground state (GS) from
ferromagnetic (FM) to anti-ferromagnetic (AFM). In Ref.~\cite{sub99}
it is proposed that the key structural modification is related to one
of the Cu--O--Cu super-exchange (SE) angles present in the system,
$\alpha_2$ in Fig.~\ref{fig1}. The change in $\alpha_2$ would cause
the corresponding SE coupling ($J_2$ in Fig.~\ref{fig2}) to switch
sign, thus transforming the GS from FM to the AFM2 spin configuration
of Fig.~\ref{fig2}. This interpretation follows the spirit of the
well-known Anderson-Goodenough-Kanemori (AGK) rules~\cite{and63},
which discuss the SE sign and strength as a function of atomic species
and configurations. However, we should note that, to the best of our
knowledge, there is no direct experimental evidence that AFM2 is the
GS of the Te-rich alloys, as the spin structure is yet to be
determined by neutron scattering measurements.

Motivated by this appealing physical picture, we decided to study the
Se$_{1-x}$Te$_x$CuO$_3$ alloys using Density Functional Theory (DFT)
and complementary tight-binding and many-body techniques. In this
Letter we report our surprising results. We find that (i) the changes
in SE angles with $x$ have negligible influence on the corresponding
couplings and (ii) what controls the magnetic interactions is the
position of the presumedly {\sl dummy} atoms Se and Te with respect to
some Cu--O--Cu SE paths. Indeed, the Se/Te atoms seem to act like a
{\sl valve}, turning the AFM SE off as they approach the Cu--O--Cu
group. Our results thus draw a picture of STCO that is much more
subtle than that proposed in Ref.~\cite{sub99}. At the same time, they
hint at new general ways of engineering magnetic couplings.

\begin{table}[b!]
\caption{Structural parameters defined in
Fig.~\protect\ref{fig1}. Values taken from
Refs.~\protect\cite{koh76,phi76}. Distances in Angstroms and angles in
degrees. Unit cell volume $\Omega$ in \AA$^3$.}
\smallskip
\begin{tabular}{lcccccccc}
\hline\hline
system & $\Omega$ & $\alpha_1$ & $d_1$ & $\tilde{d}_1$ & $\alpha_2$ &
$d_{2}/d_{2}'$ & $\tilde{d}_2$ & $d_{\rm A-O(1)}$ \\
\hline
SeCuO$_3$ & 231 & 122.4 & 2.09 & 3.66 & 127.1 & 1.92/2.52 & 3.98 & 1.75\\
TeCuO$_3$ & 245 & 123.5 & 2.06 & 3.63 & 130.5 & 1.90/2.61 & 4.11 & 1.90\\
\hline\hline
\end{tabular}
\label{table1}
\end{table}

The calculations were performed within the generalized gradient
approximation (GGA-PBE~\cite{per96}) to DFT. We primarily used the
all-electron implementation in the WIEN2k package~\cite{wien2k}, with
a mixed basis that includes augmented plane waves and local orbitals
(APW+lo). We used the LDA+U scheme to properly treat the 3$d$
electrons of Cu~\cite{ani93,lie95}. Typical cuprate values were taken
for $U$ (7.5~eV) and $J$ (1.36~eV). We also used the ultrasoft
pseudopotential~\cite{van90} implementation in the PWscf
package~\cite{pwscf}, with the LDA+U approach of Ref.~\cite{coc} and
$U$=6~eV. The calculation conditions~\cite{fn-calcs} were converged to
obtain exchange constants with an accuracy better than 1~meV. We
checked that variations of 0.5-1~eV in $U$ do not change our
qualitative results. We double-checked all our results by performing
both WIEN2k and PWscf calculations. In all cases we got full
qualitative, and reasonable quantitative, agreement.


{\sl Raw ab initio results.--} We start by considering SeCuO$_3$ (SCO)
and TeCuO$_3$ (TCO) in their experimental
structures~\cite{koh76,phi76}. Both compounds have a 20-atom unit cell
and differ only by small variations in atomic positions and lattice
constants. The structure is shown in Fig.~\ref{fig1} and the relevant
structural data is given in Table~\ref{table1}.

We describe the magnetic interactions by means of a Heisenberg
Hamiltonian $H=1/2\sum_{i,j}J_{ij}\vec{S}_i \cdot\vec{S}_j$ in which
we include the exchange constants $J_1$, $J_2$, and $J_3$ defined in
Fig.~\ref{fig2}. ($J_1$ and $J_2$ are, respectively, associated to SE
angles $\alpha_1$ and $\alpha_2$.) We compute the $J$'s by requiring
that this Hamiltonian reproduces, at a classical level, the energy
differences between the spin configurations in Fig.~\ref{fig2}
calculated from first-principles.

\begin{table}[b!]
\caption{Exchange constants of Fig.~\protect\ref{fig2} calculated for
various systems (see text). Values are given in meV and the magnetic
ground states (GS) are indicated. The results for SCO and TCO are
consistent with high-$T$ expansion fits of the susceptibility data in
Ref.~\protect\cite{sub99}. In TCO there is competition between $J_2$
and $J_3$, which probably leads to interesting spin dynamics.}
\smallskip
\begin{tabular}{lcrrr}
\hline\hline
system       & GS   &  $J_1$  & $J_2$  & $J_3$  \\
\hline
SCO               & FM   &  $-$4.4 & $-$1.3 & $-$0.8 \\
TCO               & AFM1 &     6.3 & $-$1.5 & $-$0.5 \\
SCO/TCO-st        & AFM1 &    17.7 & $-$2.3 & $-$0.6 \\
TCO/SCO-st        & FM   & $-$14.3 &    1.1 & $-$0.7 \\
\hline\hline
\end{tabular}
\label{table2}
\end{table}

Our {\sl ab initio} all-electron results for SeCuO$_3$ and TeCuO$_3$
are given in the first two lines of Table~\ref{table2}. In agreement
with experiment, we find FM and AFM ground states for SCO and TCO,
respectively. However, the calculations predict that the GS of TCO is
AFM1, and not AFM2 as proposed in Ref.~\cite{sub99}. Accordingly, it
is $J_1$, not $J_2$, the magnetic coupling that changes sign when
going from SCO to TCO. In fact, even though the change in $\alpha_1$
is around three times smaller than that in $\alpha_2$ (see
Table~\ref{table1}), $J_1$ varies by about 200\% of its value, while
$J_2$ remains almost constant. This clearly indicates that the SE
angles have little influence in the magnetic couplings of these
alloys. The last two lines of Table~\ref{table2} show the results
obtained when we consider SeCuO$_3$ in the TeCuO$_3$ structure
(denoted by ``SCO/TCO-st'') and viceversa. The results confirm that it
is the structure, and not chemical differences between Se and Te, what
determines the magnetic GS.


\begin{figure}
\begin{center}\includegraphics[width=8cm]{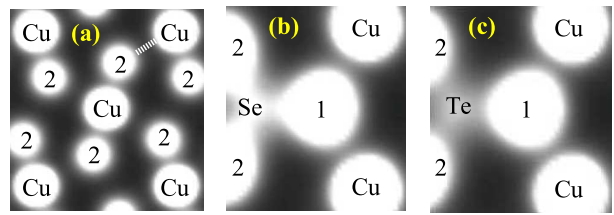}\end{center}
\caption{Calculated spin-up charge densities. Panel~a: Cu--O(2)--Cu
groups in $ac$ plane of SeCuO$_3$ (see Fig.~\protect\ref{fig1}, right
top). Dashed line marks one Cu--O(2) {\sl broken bond} (see
text). Panels~b and c: Cu--O(1)--Cu group and neighboring A cation,
for SeCuO$_3$ and TeCuO$_3$, respectively (see
Fig.~\protect\ref{fig1}, right bottom).}
\label{fig3}
\end{figure}

We identify the causes of these results by examining the electronic
densities that come out of the calculations. Figure~\ref{fig3}a shows
the spin-up density along Cu--O(2)--Cu paths in the $ac$ plane of SCO
(the TCO result is essentially the same). As it is obvious, there are
Cu--O(2) {\sl broken bonds}. This is not so surprising when one notes
that the {\sl broken-bond} distance, $d_{2}'$, is 2.52~\AA\, while the
other Cu--O(2) distance, $d_2$, is only 1.92~\AA\ (see
Table~\ref{table1}; the values for TCO are similar). The typical Cu--O
distance in cuprates is 2~\AA, suggesting that in SCO and TCO the SE
contribution to $J_2$ will be unconventional and weaker than usual. In
fact, it is questionable that the above mentioned AGK rules apply in
this case, and it seems reasonable that $J_2$ is largely independent
of $\alpha_2$.

\begin{figure}
\begin{center}\includegraphics[width=8cm]{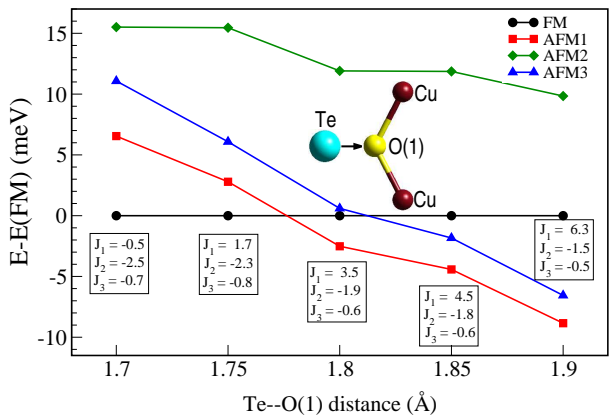}\end{center}

\vspace{-5mm}

\caption{Calculated energy of spin configurations defined in
Fig.~\protect\ref{fig2}, for TeCuO$_3$, as a function of the Te--O(1)
distance. FM configuration is taken as the zero of energy. Calculated
exchange constants are given in meV.}
\label{fig4}
\end{figure}

Figures~\ref{fig3}b and \ref{fig3}c show the spin-up charge density
along the Cu--O(1)--Cu path for SCO and TCO, respectively. As far as
the Cu--O distances are concerned (see Table~\ref{table1}), this SE
path is similar in both systems and a more conventional one. However,
there is a structural feature that makes a big difference between SCO
and TCO, namely, the position of the neighboring A cation with respect
to the O(1) atom. In SCO, $d_{\rm A-O(1)}$ is 1.75~\AA, while we have
1.90~\AA\ in TCO. Accordingly, as the density plots in
Figs.~\ref{fig3}b and \ref{fig3}c suggest, the Se$^{+4}$ cation
probably perturbs the O-2$p$ orbitals more than Te$^{+4}$ does. One
may thus hypothesize that this perturbation somehow disrupts the SE
mechanism and renders a FM $J_1$ in SCO, while regular SE results in
an AFM $J_1$ in TCO. To check this conjecture, we calculated the
magnetic interactions in TCO as a function of the Te--O(1)
distance. The results in Fig.~\ref{fig4} show that, indeed, when Te
comes close enough to O(1) (about 1.75~\AA), a FM GS results and $J_1$
switches sign. This is very strong evidence that we have identified
the structural feature that controls the magnetic ground state in
Se$_{1-x}$Te$_x$CuO$_3$ alloys.

{\sl Tight-binding analysis.--} Great insight can be gained into the
ultimate causes of these effects by discussing them in terms of the
relevant electronic interactions in the system (i.e., hoppings,
Coulomb, and exchange). To do so, we have implemented a simple and
powerful scheme to compute tight-binding (TB) Hamiltonians that
reproduce the first-principles electronic structure. In the following
we sketch the method.

Consider the atomic orbitals (AO's) of an isolated atom $\kappa$. Let
us bring those AO's into the crystal and call them $\phi^{\bf
R}_{\kappa\alpha}$, where ${\bf R}$ is a lattice vector and $\alpha$
stands for the quantum numbers $\{n,l,m\}$. The corresponding
Bloch-like wave function is $\phi^{\bf k}_{\kappa\alpha} = \sum_{\bf
R} e^{i{\bf k}{\bf R}} \phi^{\bf R}_{\kappa\alpha}$, where ${\bf k}$
is in the Brillouin zone (BZ). Let $\psi_{{\bf k}j}$ and $E_{{\bf
k}j}$ denote, respectively, the eigenstates and eigenvalues of the
Kohn-Sham Hamiltonian ($\hat{H}^{\rm KS}$) at ${\bf k}$. We define
\begin{equation}
|\widetilde{\phi}^{\bf k}_{\kappa\alpha}\rangle \; \equiv \; f \sum_j
|\psi_{{\bf k}j}\rangle \langle \psi_{{\bf k}j} |
\phi^{\bf k}_{\kappa\alpha} \rangle \; ,
\label{eq1}
\end{equation}
where $f$ is a normalization factor and the band index $j$ runs over a
set of bands that we are free to choose. We now perform the well-known
L\"owdin-Mattheiss transformation to obtain orthonormal wave functions
\begin{equation}
| \widehat{\phi}^{\bf k}_{\kappa\alpha} \rangle \; = \; \sum_j
|\psi_{{\bf k}j}\rangle \Bigl[ \sum_{\kappa'\alpha'}
(S^{\bf k})^{-1/2}_{\kappa'\alpha'\,;\kappa\alpha} \langle \psi_{{\bf k}j} |
\phi^{\bf k}_{\kappa'\alpha'}\rangle \Bigr] \; ,
\label{phihat}
\end{equation}
where $(S^{\bf k})^{-1/2}$ is derived from the overlap matrix of the
$\widetilde{\phi}^{\bf k}$'s as described in Ref.~\cite{mat70}. Then,
we can easily compute the associated Wannier functions (WF's)
$\widehat{\phi}^{\bf R}_{\kappa\alpha}$ and the parameters of the
corresponding TB Hamiltonian:
\begin{eqnarray}
&& \langle \widehat{\phi}^{\bf R}_{\kappa\alpha} | \hat{H}^{\rm KS} |
\widehat{\phi}^{{\bf R}'}_{\kappa'\alpha'} \rangle  = \nonumber \\ 
&& \frac{1}{N} \sum_{\bf k} e^{i{\bf k}({\bf R}-{\bf R}')} \sum_j
E_{{\bf k}j} (C^{\bf k}_{j;\kappa\alpha})^* C^{\bf
k}_{j;\kappa',\alpha'} \; ,
\end{eqnarray}
where the $C$'s are the bracketed coefficients in Eq.~(\ref{phihat})
and $N$ is the number of k-points in the BZ. 

If in the above sums we include a {\sl small} number of bands (e.g.,
the nominal Cu-3$d$ and O-2$p$ bands), the resulting WF's will
significantly differ from original AO's $\phi$ and incorporate the
effects of the surrounding lattice. The more additional bands we
include, the more the WF's will resemble the isolated-atom
orbitals. Note also that, by construction, our TB Hamiltonians can
reproduce the electronic band structure with arbitrary precision, the
only limitation being the spatial cutoff beyond which the hoppings are
neglected in practice~\cite{fn-wf}.

We used this scheme to construct TB Hamiltonians for the
non-spin-polarized band structures of SCO and TCO obtained from
pseudopotential calculations. We considered all valence and low-lying
conduction bands, as we were interested not only in the
Cu-3$d$--O-2$p$ couplings, but also in how those are modified by the A
cations. The resulting WF's are very close to the isolated-atom AO's.

The Cu-3$d$--O-2$p$ hoppings along the $J_2$ SE path are very
asymmetric, as the two Cu--O(2) distances differ greatly. For the {\sl
broken-bond} pair, we obtain a maximum hopping of 0.35~eV, while we
get 0.85~eV for the other Cu--O(2) pair. (These are results for
SCO. The situation is similar in TCO.) This quantitatively confirms
that the SE contribution to $J_2$ will be relatively small.

Regarding the other SE path, the two Cu--O(1) pairs are equivalent by
symmetry, and the maximum Cu-3$d$--O-2$p$ hopping is 0.77~eV in SCO
and 0.87~eV in TCO. On the other hand, the maximum Se-4$s$--O(1)-2$p$
hopping is 4.51~eV in SCO, while for Te-5$s$--O(1)-2$p$ in TCO we get
3.65~eV. These results indicate a relatively strong Se-4$s$--O-2$p$
interaction in SCO, which results in 3$d$--2$p$ hoppings about 10$\%$
smaller than the corresponding ones in TCO. One can thus infer that
the SE contribution to $J_1$ will be smaller in SCO than in TCO.


{\sl Many-body analysis.--} Our first-principles TB Hamiltonians can
be supplemented with Coulomb and exchange terms to obtain realistic
models of the electronic interactions. The resulting Hubbard-like
Hamiltonians make it possible to isolate the various contributions to
the magnetic couplings (direct-exchange, SE), and thus identify the
mechanisms behind them. Here we focus on the description of the
low-energy spin excitations in particular Cu--O--Cu groups, where the
magnetic interaction is described by a single exchange constant $J$
(this will be $J_1$ or $J_2$ if we consider, respectively,
Cu--O(1)--Cu or Cu--O(2)--Cu). The many-body states are constructed by
distributing two holes among the Cu-3$d$ and O-2$p$ orbitals. We
diagonalize the resulting Hamiltonian matrix and compute $J$ from the
energy gap between the lowest-lying singlet and triplet states.

We added the following terms to our first-principles TB Hamiltonians:
on-site Coulomb interactions for both Cu-3$d$ and O-2$p$ electrons,
on-site exchange for the O-2$p$ electrons (which implements Hund's
first rule), and inter-site exchange between Cu-3$d$ electrons (which
is well-known to favor FM interactions). The resulting Hamiltonians
are a simplification of that in Eq.~(4) of Ref.~\cite{tan95}. We
retained the above terms partly guided by numerical evidence that they
have the biggest effect in the $J$'s. We treated all Cu-3$d$ and all
O-2$p$ orbitals equally, so that the model has only four parameters,
namely, Coulomb $U_d$ and $U_p$, and exchange $K_p$ and $K_{dd}$.

By choosing reasonable values of these parameters, we were able to
obtain $J_1$ and $J_2$ exchange constants in qualitative agreement
with the first-principles results for both SCO and TCO. We employed
parameters in the following ranges: $U_d$=8-9~eV, $U_p$=5-6~eV,
$K_p$=1-2~eV, and $K_{dd}$=8.5-12.5~meV. We imposed the constraint
that $K_{dd}$ is consistent with the Cu--Cu distance, i.e., we used
larger values for smaller distances. Our analysis led us to two main
conclusions. (i) In all the cases, the SE couplings {\sl per se}
result in AFM interactions. The inter-site exchange $K_{dd}$ is
necessary to obtain FM $J$'s. (ii) The structural changes tune the
magnetic couplings via their effect on the magnitude of the
Cu-3$d$--O-2$p$ hoppings. Smaller hoppings result in a smaller SE
singlet-triplet splitting and, thus, make it easier for the $K_{dd}$
interaction to turn the coupling FM. Given the drastic approximations
underlying our Hubbard-like Hamiltonians, and the fact that the $U$
and $K$ parameters were not calculated {\sl ab initio}, these
conclusions should be taken with caution. Nevertheless, they are
consistent with our body of results and make physical sense.

In summary, we have studied the Se$_{1-x}$Te$_x$CuO$_3$ alloys in
which the magnetic interactions are controlled by
composition-dependent structural changes. We employed three
complementary approaches, i.e., LDA+U, tight-binding, and exact
diagonalization of effective Hamiltonians for finite clusters, which
allowed us to study the magnetic interactions in great detail. We find
that the key structural feature is the position of the supposedly {\sl
dummy} atoms Se and Te relative to Cu--O--Cu SE paths. As the Se/Te
atom approaches the Cu--O--Cu group, it acts like a magnetic {\sl
valve} and reduces the SE contribution to the magnetic coupling. We
find SE favors anti-ferromagnetism and, thus, this decrease allows
direct exchange between Cu-3$d$ electrons to render a ferromagnetic
interaction. This is quite a surprising mechanism, as a more
conventional one related to the SE angles might have been expected
and, in fact, was proposed in the literature~\cite{sub99}. On the
other hand, this effect could well lead to novel ways of engineering
magnetic couplings in solids. Our work shows that complex magnetic
interactions may underlay seemingly simple phenomena, and highlights
the usefulness of {\sl ab initio} studies.

It would be interesting to confirm experimentally the magnetic ground
state of TeCuO$_3$ that we predict. Also, we expect that application
of external pressure on TeCuO$_3$ could allow to tune the magnetic
couplings in the way shown here. Finally, we are currently
investigating other perovskites (e.g., MTiO$_3$, with M=Y, La) to
check whether similar effects occur.

We thank Art Ramirez, Collin Broholm, and David Vanderbilt for useful
comments.

\end{document}